# Tunneling magnetoresistance in altermagnetic $RuO_2$-based magnetic tunnel junctions

Seunghyeon Noh[1†], Gye-Hyeon Kim[2†], Jiyeon Lee[1], Hyeonjung Jung[1], Uihyeon Seo[2], Gimok So[2], Jaebyeong Lee[1], Seunghyun Lee[1], Miju Park[2], Seungmin Yang[2], Yoon Seok Oh[2], Hosub Jin[2], Changhee Sohn[2*], Jung-Woo Yoo[1,5*]

1. Department of Materials and Science and Engineering, Ulsan National Institute of Science and Technology, Ulsan 44919, Republic of Korea

2. Department of Physics, Ulsan National Institute of Science and Technology, Ulsan 44919, Republic of Korea.

†S.N. and G.-H. K. contributed equally to this work.

*Corresponding author email: jwyoo@unist.ac.kr; chshon@unist.ac.kr

**Abstract**

Altermagnets exhibit characteristics akin to antiferromagnets, with spin-split anisotropic bands in momentum space. $RuO_2$ has been considered as a prototype altermagnet; however, recent reports have questioned altermagnetic ground state in this material. In this study, we provide direct experimental evidence of altermagnetic characteristics in $RuO_2$ films by demonstrating spin-dependent tunneling magnetoresistance (TMR) in $RuO_2$-based magnetic tunnel junctions. Our results show the spin-splitted anisotropic band structure of $RuO_2$, with the observed TMR determined by the direction of the Néel vector of $RuO_2$. These results reflect the altermagnetic nature of $RuO_2$ and highlight its potential for spintronic applications, leveraging the combined strengths of ferromagnetic and antiferromagnetic systems.

# Introduction

Developing new platforms for memory and logic devices is imperative to address critical challenges associated with traditional Si-based devices, including continuous downscaling, excessive Joule heating, and the need for low-power operation. Harnessing collective state variables offers a promising pathway to achieve further miniaturization, improved energy efficiency, and inherent non-volatility. Spin-based devices, which utilize the collective state of ferromagnetism with its non-volatile properties, have demonstrated significant potential and versatility for next-generation technological applications. Antiferromagnetic (AFM) spintronics has recently emerged as a promising subfield of spintronics, leveraging the AFM order parameter, known as the Néel vector, as a state variable[1,2]. Antiferromagnets offer distinct advantages, including immunity to magnetic perturbations, absence of stray fields, and ultrafast spin dynamics, making them promising candidates for spintronic applications[3]. These properties position antiferromagnets as potential replacements for ferromagnets, offering enhanced switching speeds and storage densities. However, to fully exploit their potential, efficient electrical control and reliable detection of the Néel vector are critical.

The emergence of altermagnets offers a transformative opportunity in this context. Altermagnets combine the advantages of antiferromagnetic order with spin-split anisotropic bands in momentum space, enabling efficient electrical control and detection of the Néel vector. As a third fundamental magnetic phase, altermagnets possess zero net magnetization, like antiferromagnets, but break spin degeneracy in reciprocal space[4-6], enabling efficient spin current generation and detection. Theoretical studies have identified a class of collinear altermagnetic compounds based on symmetry considerations[4]. Among these, $RuO_2$ has been proposed as a prototype altermagnet, with predictions of a large spin-splitting of up to 1.4 eV[7-9].

Early studies employing resonant x-ray scattering[10] and neutron diffraction[11] have identified antiferromagnetic (AFM) order in $RuO_2$. Transport experiments have subsequently revealed unconventional phenomena, including anomalous Hall effects and Edelstein-like spin splitting effects which align closely with theoretical predictions based on the altermagnetic framework[12-26]. These findings suggest that $RuO_2$ exhibits unique electronic and magnetic properties, positioning it as a promising candidate for further exploration within the context of altermagnetism. Additionally, theoretical studies have predicted giant tunneling magnetoresistance (TMR) in (110)-oriented $RuO_2$-based magnetic tunnel junctions (MTJs), such as $RuO_2/TiO_2/RuO_2$ [27,28] and $RuO_2/TiO_2/CrO_2$ [29,30], further underscoring its potential.

However, recent investigations using muon-spin rotation ($\mu$-μSR) studies, a technique highly sensitive to detecting local magnetic moments, have revealed exceptionally small magnetically ordered moments in $RuO_2$ in both single-crystal and thin-film samples[31,32]. Furthermore, angle-resolved photoemission spectroscopy (ARPES) studies have failed to observe the expected band splitting attributed to altermagnetism [33]. These findings challenge the previously proposed altermagnetic ground state in rutile $RuO_2$ [7,34] and highlight the need for experimental evidence that directly demonstrates altermagnetic characteristics. A potential approach to resolve this controversy is to experimentally confirm Néel vector-dependent TMR in $RuO_2$-based MTJs. TMR could serve as a direct probe of spin-dependent electronic structures and provide unambiguous evidence of spin-polarized transport, critical for establishing the utility of $RuO_2$ in spintronic applications.

In this work, we present Néel vector-dependent TMR in $RuO_2$-based magnetic tunnel junctions. Our findings reflect the spin-splitted anisotropic band structure of $RuO_2$, supporting its altermagnetic nature and potential for spintronic devices. The observed TMR is found to be strongly dependent on the Néel vector orientation and the crystal facet of $RuO_2$, with a sign reversal upon the inversion of the Néel vector. The observed TMR reaches a maximum magnitude of approximately ~5% at 10 K, gradually diminishing with increasing temperature and vanishing above 50 K. These results provide crucial evidence for the altermagnetic properties of $RuO_2$ film, highlighting its potential as a candidate for spintronic devices by leveraging the combined advantages of ferromagnetic and antiferromagnetic systems.

Ruthenium dioxide ($RuO_2$) is a rutile-structured oxide belonging to the tetragonal $P4_2/mnm$ space group, characterized by Ru atoms positioned at the center of stretched oxygen octahedrons as shown in Fig. 1a. Each Ru atom in different sublattices is surrounded by oxygen octahedra oriented in distinct directions, rotated by 90°. These sublattice rotations in real space lead to elliptical electronic bands for up and down spins, with 90° rotation in momentum space, producing a d-wave-like spin-split band structure. Fig. 1b illustrates that this unique anisotropic spin bands results in a balanced spin population within the (001) 2D Brillouin zone, but generates a polarized spin population in the (110) 2D Brillouin zone when the Néel vector is oriented along the [001] direction[35]. Due to the *k*-dependent spin polarization, the giant TMR was predicted in symmetric $RuO_2/TiO_2/RuO_2$ MTJ[27,28]. In this case, tuning the orientation of the Néel vector could be achieved through the spin-orbit-torque or exchange bias, which requires further extensive works. Conventional ferromagnetic materials can serve as a counter electrode to probe Néel vector dependent TMR in $RuO_2$–based systems[29,30], as depicted in Fig. 1c.

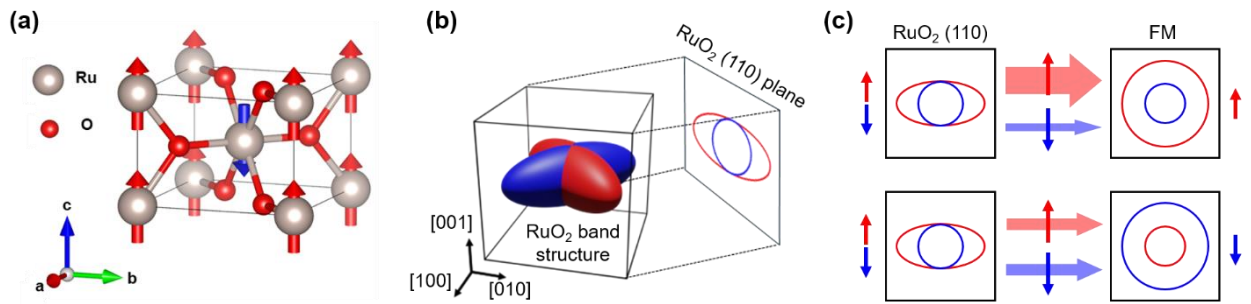

Fig. 1. Schematic illustration for the crystal and spin-splitted band structure of $RuO_2$ and projected TMR in $RuO_2$-based MTJ. (a) The crystal structure of $RuO_2$ is depicted with spin orientations indicated by red and blue arrows, representing its collinear antiferromagnetic ordering. (b) Schematic illustration of Fermi surface of $RuO_2$ displaying compensated spin polarization in the (001) plane and uncompensated spin polarization in the (110) plane. (c) Schematic illustration of TMR in $RuO_2(110)/TiO_2$/FM MTJ. Red and blue curves represent up- and down-spin Fermi surfaces, respectively. Large arrow indicates tunneling current and its width is proportional to the tunneling probability, reflecting the spin-dependent tunneling across the junction.

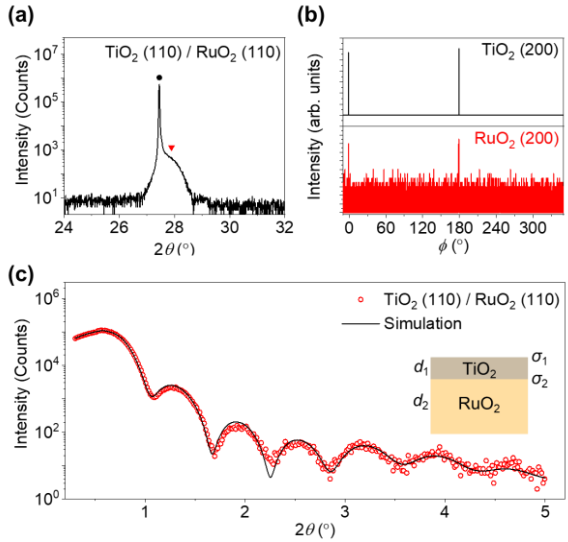

FIG. 2. (a) X-ray diffraction (XRD) $\theta$–$2\theta$ scan of the TiO$_2$/RuO$_2$ bilayer heterostructure. The black circles and red triangles correspond to the TiO$_2$ (110) and RuO$_2$ (110) diffraction peaks, respectively. (b) XRD $\varphi$-scan of the TiO$_2$ (200) plane (top) and RuO$_2$ (200) plane (bottom), demonstrating the epitaxial relationship. (c) X-ray reflectivity (XRR) measurement of the TiO$_2$/RuO$_2$ bilayer structure (red circles) with the best-fit simulation (black line). Inset: Model structure and structural parameters used for the simulation.

We synthesized TiO$_2$/RuO$_2$ bilayer heterostructure via pulsed laser deposition (PLD) and characterized their structural properties using X-ray diffraction (XRD) and X-ray reflectivity (XRR) (see Supplemental Material S1 [36]). Figure 2(a) presents the XRD $\theta$–$2\theta$ scan of RuO$_2$ (110), showing a well-defined RuO$_2$ (110) peak near 28°, confirming single-crystalline heterostructure. In Fig. 2b, XRD $\varphi$-scan of $RuO_2$ (200) and $TiO_2$ (200) plane are presented for confirming the epitaxial growth of $RuO_2$ and $TiO_2$ layer and the in-plane crystal orientation can be determined by the $\varphi$-scan. XRR measurements (Fig. 2c) were performed to determine the thickness and roughness of each layer in theheterostructure. By fitting the oscillation curve, the structural parameters, including the layer thicknesses ($d_1$, $d_2$) and interface roughnesses ($\sigma_1$, $\sigma_2$), were extracted with high precision. This analysis confirms that all deposited heterostructures exhibit exceptionally flat interfaces with minimal roughness. However, some regions of the heterostructure surface contain particles or residual islands due to the $RuO_2$ growth mode [37], which can impact the interface quality and affect the tunnel barrier uniformity and TMR ratio [see Supplementary Materials S2]

To demonstrate the Néel vector dependent TMR, we fabricated $RuO_2$/$TiO_2$/CoFeB MTJs using photolithography, ion miller, and magnetron sputtering. Starting from pre-deposited $RuO_2$/$TiO_2$ layers, 200 μm-wide $RuO_2$ channel was defined by Ar ion milling and 10 μm-diameter MTJ patterns were isolated using $Al_2O_3$ buffer layer. The ferromagnetic electrode was deposited as a 40 nm CoFeB layer via UHV sputtering, followed by a 2 nm-thick Al capping layer to prevent oxidation. Two devices are mainly presented in the main text: Device A, comprising $RuO_2$(10.3 nm)/$TiO_2$(2 nm)/CoFeB(40 nm), and device B, comprising $RuO_2$(11.2 nm)/$TiO_2$(2.3 nm)/CoFeB (40 nm).

Fig. 3 displays measurement configuration and the observed TMR in our $RuO_2$/$TiO_2$/CoFeB MTJ device A. Here, we define TMR as TMR = 100×($R(+H) - R(-H)$)/$R(+H)$, where $H$ =

200 Oe. TMR was measured by sweeping magnetic field both along the [001] Néel vector direction and the [1$\bar{1}$0] direction, which is perpendicular to the Néel vector. Initially, the magnetoresistance (MR) of the MTJ device A was measured at 10 K after cooling the sample without applying an external magnetic field, ensuring that $RuO_2$ retained a domain structure with unoriented Néel vectors. Fig. 3b displays the observed MR for magnetic field sweeps along the [001] direction. Under the zero-field cooling conditions, the device exhibited very weak signature of spin-dependent tunneling, consistent with the unaligned Néel vector. To align the Néel vector, the device was warmed to 370 K and subsequently cooled down to 10 K under an applied magnetic field of 1 T. Fig. 3c and d show the TMR measurements for the field-cooled device under magnetic field sweeps along the [001] and [1$\bar{1}$0] directions, respectively. For the [001] field sweep, the TMR exhibited a clear hysteresis loop characteristic of CoFeB, confirming spin-dependent tunneling into $RuO_2$. In contrast, for the [1$\bar{1}$0] field sweep, the MRs measured at +200 Oe and −200 Oe are nearly identical. However, a significant TMR emerged within the hysteresis loop of CoFeB. Near the coercive field, the magnetization of CoFeB momentarily aligns with the Néel vector due to the rotation of magnetic moments. Remarkably, the maximum TMR observed under these conditions closely approaches the magnitude of the TMR measured during the [001] field sweep. Fig. 3e shows angle-dependence of MR. The blue symbols represent the MR data recorded while rotating the magnetic field of +200 Oe in the plane, whereas red symbols correspond to the MR data recorded under a rotating magnetic field of -200 Oe. The data reveals a distinct angular dependence of MR. Fig. 3f shows the estimated angle dependent TMR, which follows a characteristic ~ $\cos\theta$ dependence. This behavior is consistent with TMR associated with Néel vector, which does not reorient to the small applied magnetic field. We have fabricated a number of MTJ devices to test TMR. Results displays the measured TMR ratios were correlated with *RA* product (resistance×area) as shown in Supplemental Material S3 [36].

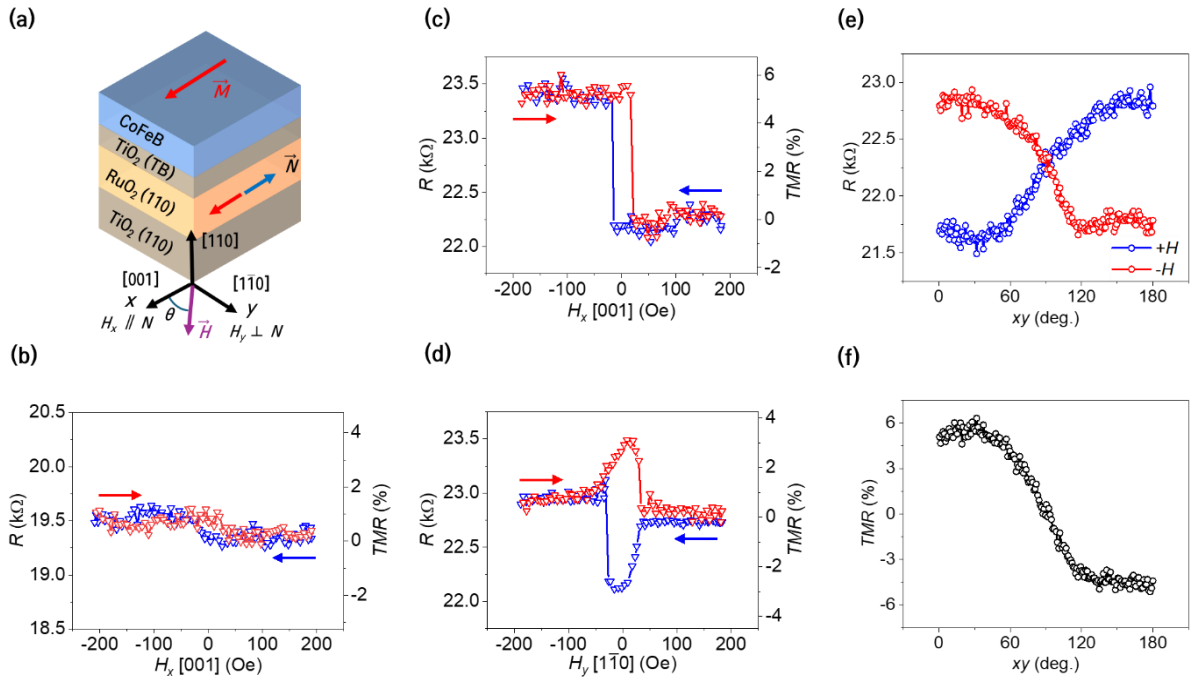

Fig. 3. Néel vector-dependent TMR in $RuO_2$/$TiO_2$/CoFeB MTJ (Device A). (a) Schematic view of $RuO_2$/$TiO_2$/CoFeB MTJ and magnetic field direction used for TMR measurements. (b) The MR curves recorded at 10 K during magnetic field sweeps along the [001] direction without prior alignment of the Néel vector. The red (blue) up-triangles (down-triangles) indicates data collected with increasing (decreasing) magnetic field. (c, d) MR curves measured at 10 K during magnetic field sweeps along the [001] and [1$\bar{1}$0] direction after aligning the Néel vector via field-cooling at 1 T along the [001] direction from 370 K. (e) The angle dependent MR curves measured at 10 K under +200 Oe (Blue) and −200 Oe (Red) in *xy* plane. (f) The estimated TMR as a function of magnetic field angle in *xy* plane.

To further investigate the effect of Néel vector inversion, we annealed the sample (device B) under the large negative magnetic fields from 370 K. Fig. 4a exhibits the measured TMR at 10 K after field-cooling of −8T. The observed TMR shows inverted hysteresis loops compared to the results of Fig. 3c. In order to confirm the effect of Néel vector inversion, the device B was annealed again with large positive magnetic fields (+8T). As shown in Fig. 4b, the measured TMR returns to its original shape for the [001] Néel vector, similar to the case of Fig. 3c. We note that this reversible TMR effect in response to the Néel vector orientation was consistently observed in another device (See Supplemental Material S4 [36]). The temperature dependence of TMR in our devices were also investigated as shown in Supplemental Material S5 [36]. The measured TMR in our devices typically become negligible above 50 K. However, improved film and interface qualities as well as device fabrication should significantly enhance the value of TMR. Nonetheless, our results unequivocally confirm spin-dependent tunneling into $RuO_2$ and highlight its unique spin-split anisotropic band structure, reinforcing its potential for spintronic applications.

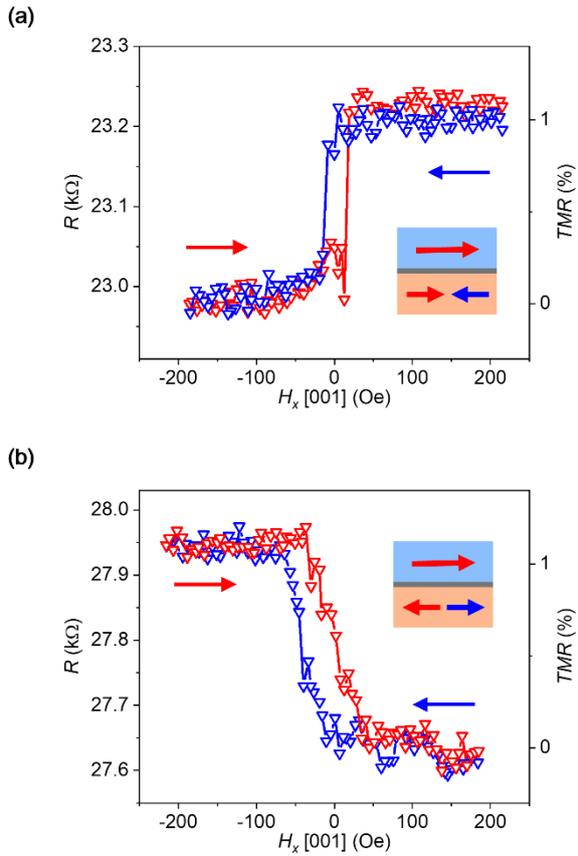

Fig. 4. The Néel vector orientation-dependent TMR in $RuO_2$/$TiO_2$/CoFeB MTJ. MR curves for device B measured at 10 K along the [001] direction. Measurements were performed after field-cooling with −8 T (a) and +8 T (b), respectively.

In this study, we demonstrated the Néel vector-dependent TMR in $RuO_2$-based magnetic tunnel junctions (MTJs), providing experimental evidence of the altermagnetic properties of $RuO_2$ film. The observed TMR strongly correlates with the orientation of the Néel vector and reflects the unique spin-split anisotropic band structure of $RuO_2$, a hallmark of its altermagnetic nature. Additionally, the inversion of the Néel vector through thermal annealing under magnetic fields further corroborates the spin-dependent tunneling behavior and highlights the tunability of this material. Our findings establish RuO$_2$ as a promising candidate for next-generation spintronic devices, combining the advantages of antiferromagnetic stability and ferromagnetic spin-split properties. The temperature-dependent TMR measurements, revealing diminishing effects above 50 K, suggest the need for further optimization to achieve room-temperature performance. These results lay a strong foundation for the development of RuO$_2$-based spintronic applications, such as high-density memory devices and ultrafast logic components, paving the way for further exploration of altermagnetic materials in functional technologies.

## Acknowledgements

This work was supported by the National Research Foundation of Korea (NRF) grant funded by the Korea government (RS-2024-00487988).

**Supplementary Materials**

**Tunneling magnetoresistance in altermagnetic RuO$_2$-based magnetic tunnel junctions**

Seunghyeon Noh[1†], Gye-Hyeon Kim[2†], Jiyeon Lee[1], Hyeonjung Jung[1], Uihyeon Seo[2], Gimok So[2], Jaebyeong Lee[1], Seunghyun Lee[1], Miju Park[2], Seungmin Yang[2], Yoon Seok Oh[2], Hosub Jin[2], Changhee Sohn[2*], Jung-Woo Yoo[1,5*]

1. Department of Materials and Science and Engineering, Ulsan National Institute of Science and Technology, Ulsan 44919, Republic of Korea.

2. Department of Physics, Ulsan National Institute of Science and Technology, Ulsan 44919, Republic of Korea.

†S.N. and G.-H. K. contributed equally to this work.

*Corresponding author email: jwyoo@unist.ac.kr; chshon@unist.ac.kr

## S1. RuO$_2$ thin film and TiO$_2$/RuO$_2$ heterostructure synthesis and characterization

The RuO$_2$ thin film and TiO$_2$/RuO$_2$ heterostructures were synthesized by pulsed laser deposition. The rutile TiO$_2$ (110) substrate (CrysTec GmbH) and the sintered target of RuO$_2$ (TOSHIMA Manufacturing Co., ltd.) are used for epitaxial RuO$_2$ thin film. the optimized growth condition for RuO$_2$ thin film were as follows: a substrate temperature of $T = 350$ °C, an oxygen partial pressure of $P = 100$ mTorr, and a laser fluence of $F = 0.75$ J/cm$^2$ with 10 Hz laser repetition. As a result of the low laser fluence, 12,000 pulses were required to achieve the desired thickness.

For the TiO$_2$/RuO$_2$ heterostructure, the TiO$_2$ layer was deposited *in situ* immediately after the RuO$_2$ layer. The TiO$_2$ deposition was carried out under the same substrate temperature and oxygen partial pressure as the RuO$_2$ layer, but with an increased laser fluence of 1.5 J/cm². Structural characterization of all RuO$_2$ thin films and RuO$_2$/TiO$_2$ heterostructures was performed using X-ray diffraction (XRD).

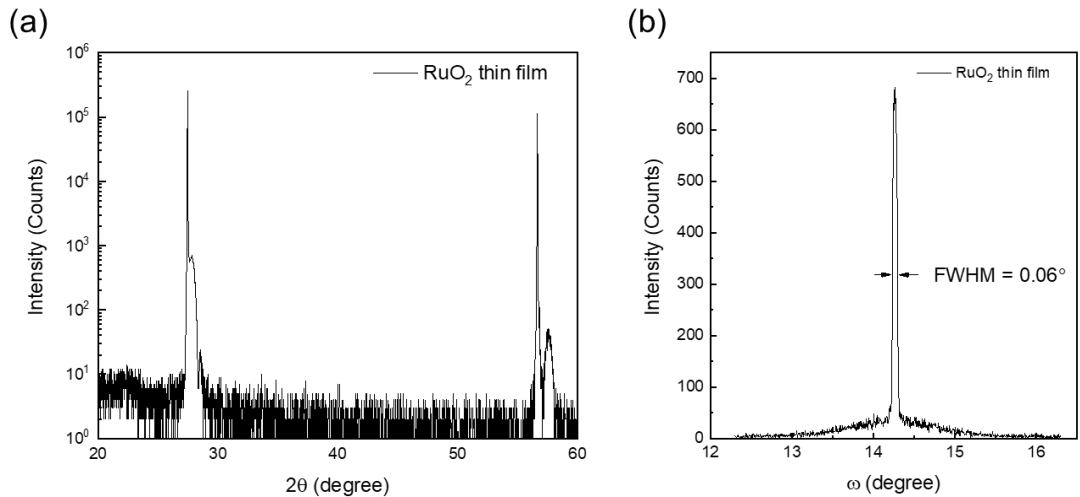

FIG. S1. (a) X-ray diffraction $\theta$-$2\theta$ scan of RuO$_2$ (110) on TiO$_2$ (110) substrate. Well-defined RuO$_2$ (110) and (220) peak is measured with thickness fringe. (b) Rocking curve of RuO$_2$ (110)

peak. The full width of half maximum of the Rocking peak is 0.06˚, confirming good crystallinity of $RuO_2$ thin film.

## S2. Interface study using atomic force microscopy

For examining the surface condition, the atomic force microscopy measurement is conducted. As shown in Fig. S2, when the $RuO_2$ layer was deposited, most surfaces have very small roughness, but some particles and/or residual islands of approximately 20 nm are often observed on the surface. Even though the $TiO_2$ tunnel barrier was deposited, the island structures on the surface still remain, which can affect the device quality and consequently influence the TMR ratio.

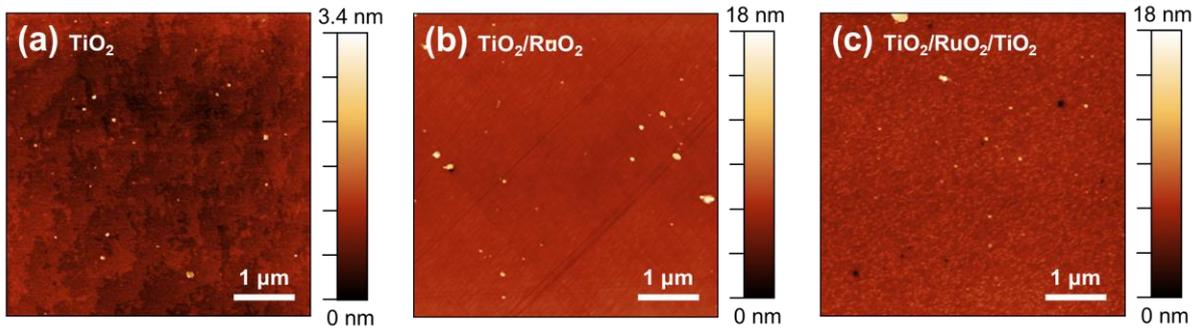

FIG. S2. Atomic force microscopy of (a) $TiO_2$ (110) substrate, (b) $RuO_2$ (110) thin film on $TiO_2$ (110) substrate, and (c) $TiO_2$ (110)/$RuO_2$ (110) bilayer heterostructure on TiO2 (110) substrate.

## S3. Resistance-area product (*RA*) dependence of TMR efficiency

We fabricated a number of MTJ samples to demonstrate the Neel vector dependent TMR effects. The fabricated $RuO_2$/$TiO_2$/CoFeB MTJs had diameters ranging from 3 to 20 μm, with 10 μm being the optimal size. We observed a significantly enhanced TMR effect at specific range of tunnel-barrier resistances. The TMR ratios as a function of the resistance-area (*RA*) product are presented in Fig. S3, showing a normal distribution trend. However, a few samples exhibit exceptionally high TMR ratios due to variations in tunnel barrier uniformity or interfacial quality.

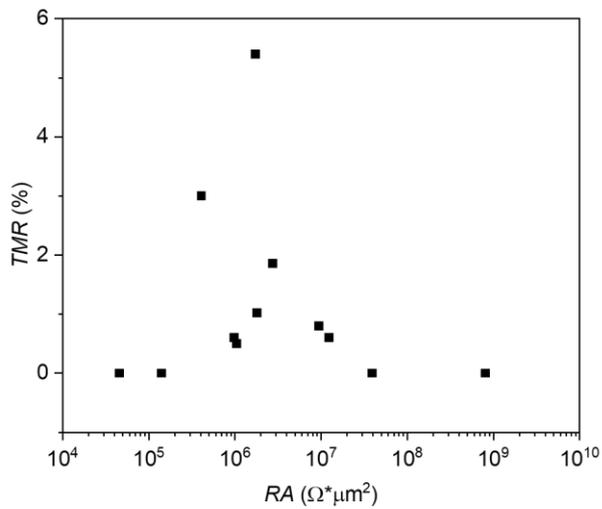

FIG. S3. The measured TMR as a function of *RA* (Resistance × area) of the devices. The size of $RuO_2$/$TiO_2$/CoFeB MTJs varies in diameters of 3 - 20 μm. The TMR data were recorded at 10 K.

**S4. Reproducibility of TMR for the Néel vector reorientation**

The Néel vector reorientation of altermagnet RuO$_2$ is challenging. In previous studies[1,2], the anomalous Hall effect in RuO$_2$ was reported under a 50 T magnetic field sweep, and Néel vector reorientation was observed through field annealing at 473 K under 8 kOe. In our approach, Néel vector reorientation was achieved through field cooling (FC). We cooled the devices from 370 K to 10 K under a high magnetic field (8 T) in a Physical Property Measurement System (PPMS) chamber. The device B (Fig. 4 in main text) provides clear evidence of the Néel vector reorientation through an inverted TMR effect. Another device C, RuO$_2$ (12.3 nm) / TiO$_2$ (2.5 nm) / CoFeB (40 nm), also exhibits the same phenomenon, confirming the reproducibility of the Néel vector reorientation. Fig. S4 (a) shows the TMR effect of device C at 10 K under −8 T field-cooled (FC) condition, while Fig. S4 (b) represents the +8 T FC condition, revealing an inverted shape of TMR curves. Although the TMR ratio is lower and noisier, the data clearly indicate an inverted TMR effect due to FC orientation, demonstrating Néel vector reorientation. These results validate the reproducibility of reversible TMR by the Néel vector reorientation.

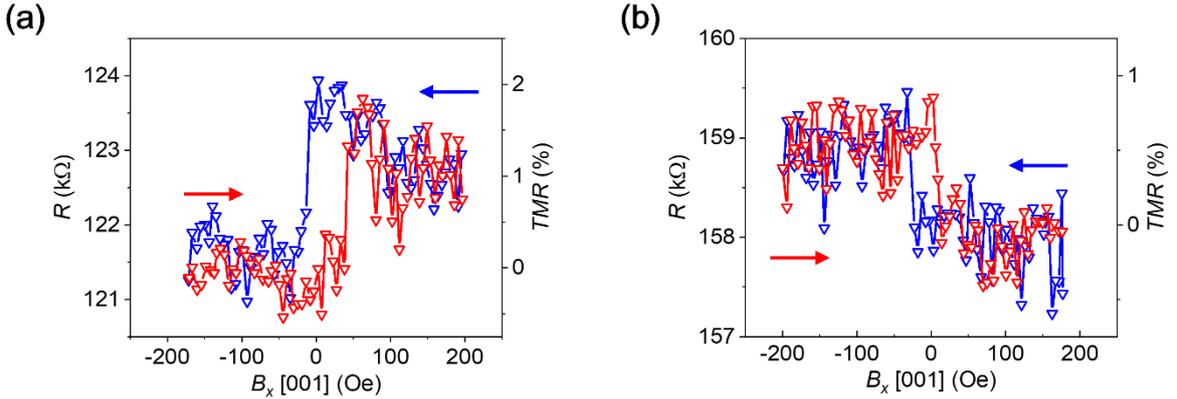

FIG. 4. Néel vector reorientation in another RuO$_2$/TiO$_2$/CoFeB MTJ (Device C). The TMR effects measured at 10 K along the [001] direction. Measurements were performed after field-cooling with −8 T (a) and +8 T (b), respectively. The measured TMR curve was inversed by Néel vector reorientation.

## S5. Temperature dependence and reproducibility of TMR in MTJs

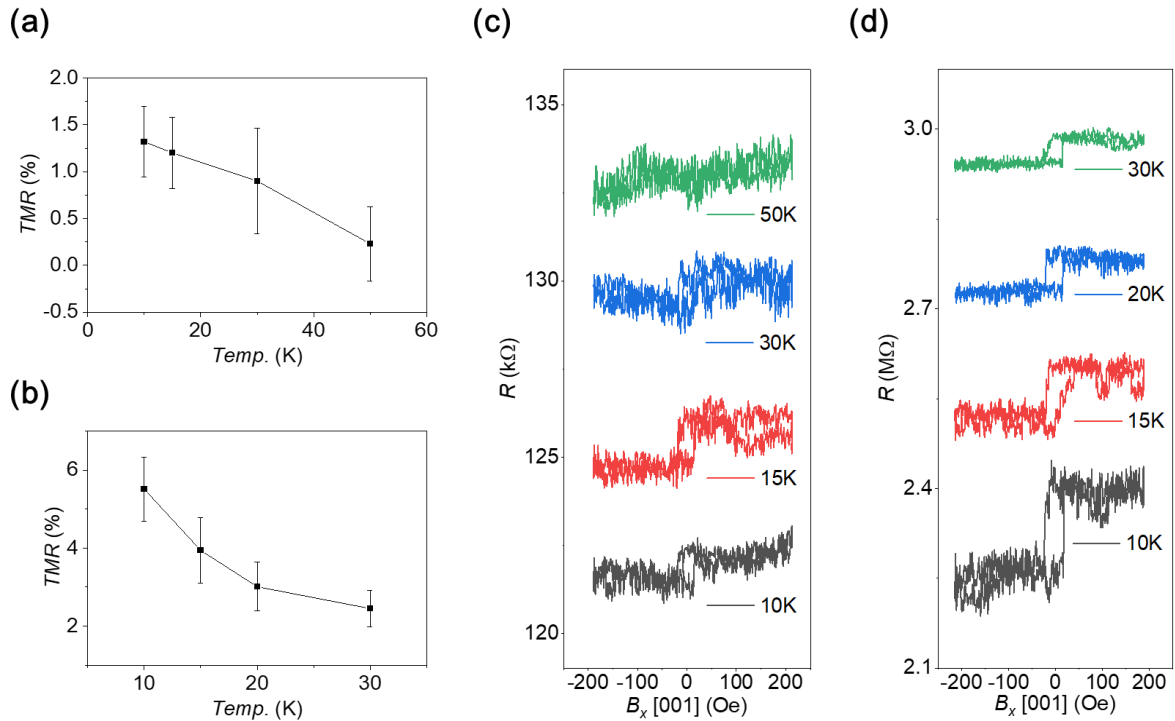

FIG. S5. Temperature dependence of TMR effect in two $RuO_2/TiO_2/CoFeB$ MTJ devices. (a, b) TMR ratio as a function of temperature for device C, and device D. (c, d) TMR curves measured at different temperatures along the [001] direction in device C and device D, respectively. Device D is comprised of $RuO_2$ (12.1 nm) / $TiO_2$ (4.4 nm) / CoFeB (40 nm). Both devices exhibit negligible signals above 50 K.